\def\beg{\begin{equation}}
\def\eeq{\end{equation}}
\documentstyle[12pt]{article}
\begin{document}
\begin{center}
{\Large{\bf Doubling of Landau levels in 
$GaAs/Al_xGa_{1-x}As$ heterostructures.}}
\vskip0.55cm
{\bf Keshav N. Shrivastava}
\vskip0.25cm
{\it School of Physics, University of Hyderabad,\\
Hyderabad  500046, India}
\end{center}

Usually the Landau levels are given by 
$(n+{1\over 2})\hbar\omega_c$ with $\hbar\omega_c=eB/mc$. There are interactions which push the electrons from ${\it l}=0$ to ${\it l}\ne 0 $ so that there is doubling in the Landau levels with $\hbar\omega_c={1\over 2}g_{\pm}eB/mc$ where ${1\over 2}g_{\pm}=({\it l}+{1\over 2}\pm s)/(2{\it l }+1)$. For $s=\pm 1/2$, two distinct values of $\omega_c$ appear whereas only one seems to be known in the literature. This is because only one value, $s=1/2$ is populated. When
microwaves of a suitable frequency such as 13.1 GHz at a field of about 2.5 Tesla, are turned on, s= -1/2 also gets populated. For singlets s=0, ${1\over 2}g_{\pm}=1/2$ so that $\omega_c={1\over 2}(eB/mc)$. This cuts the cyclotron frequency to half and doubles the number of oscillations in the resistivity as a function of magnetic field. This theory of doubling of the oscillations agrees with the experimental data in GaAs/AlGaAs. 
\vskip1.0cm
Corresponding author: keshav@mailaps.org\\
Fax: +91-40-2301 0145.Phone: 2301 0811.
\vskip1.0cm

\noindent {\bf 1.~ Introduction}

     Recently, Mani et al[1] have reported that at certain values of the the magnetic field, the resistance is found to become zero. This zero-resistance state has been of considerable interest and has been discussed extensively[2-4]. We note that this zero-resistance state is found at high magnetic fields whereas at low magnetic fields another phenomenon  which may be called ``doubling of levels" occurs. This phenomenon can be understood by the same theory used to understand high magnetic fields.

      In this note, we show that ``doubling of levels" occurs and hence the number of oscillations in the resistivity as a function of magnetic field doubles when both spin states, $\pm 1/2$ get populated. The $s$=0 state has energy levels with separation given by half of the cyclotron frequency. This factor of 1/2 is important when cyclotron frequency is used to measure the magnetic field. Thus there are two types of doubling  of levels, one due to two signs of $s=\pm 1/2$ and another type due to $s$=0. This $s$=0 type doubling is present also in the experimental data indicating the condensation of $s$=0 state.

\noindent{\bf 2.~~Theory}

In a two dimensional electron system, when a magnetic field is applied normal to the surface, with zero orbital angular momentum such as for electrons in a conduction band, the electron energies become similar to that of a harmonic oscillator, $E_n=(n+{1\over 2})\hbar\omega_c$ where
$\omega_c$ is the cyclotron frequency, $\omega_c=eB/mc$. Here $e$ is the charge of the electron, B the magnetic field, $m$ the electron mass and $c$ the velocity of light. These harmonic oscillator type levels are called ``Landau levels". The response function of such a system varies inversely as the magnetic field. The evacuation of states occurs as,
$(n+{1\over 2})= mc\epsilon_F/(eB)$. At small values of the magnetic field, a large $n$ occurs. As the value of $B$ increases, smaller and smaller values of $n$ cross the Fermi level but there is only one value of $\omega_c$. It has been reported that both the signs of spin may be considered, i.e., $j={\it l}\pm s$. In this case, the cyclotron frequency becomes[5-10],
\beg
\omega_c= {1\over 2}g_{\pm}{eB\over mc}
\eeq
where
\beg
{1\over 2}g_{\pm}= {{\it l}+{1\over 2}\pm s\over 2{\it l}+1}
\eeq
so that there are two cyclotron frequencies instead of the well known single value. When ${\it l}$=0, $s$=1/2, ${1\over 2}g_{\pm}$=1, so that well known value is recovered. When ${\it l}$=0, $s$= - 1/2, ${1\over 2}g_{\pm}$=0, the cyclotron frequency becomes zero. The zero-frequency solution is usually associated with charge-density waves or soft modes. The $S$=0 state is superconducting and the charge per particle is 1/2. Two quasiparticles are needed to make one electron. Obviously, four such particles are needed to make one Cooper pair. When electron clusters are considered, there can occur different values of ${\it l}$. As the value of ${\it l}$ is varied, the effective charge may be taken as ${1\over 2}g_{\pm}e$. For various values of ${\it l}$ and $s$ all of the predicted values match with those measured. The Zeeman energy from the above expression may be written as,
\beg
E_z= {{\it l}+{1\over 2}\over 2{\it l}+1}\mu_BH.S \pm {S\over 2{\it l}+1}\mu_BH.S
\eeq
For ${\it l}$=0, $S$=+ 1/2, $g_{\pm}$=2. When the magnetic field of the electromagnetic field of the electron is taken into account, this value becomes 2.0023. The effect of the spin-orbit interaction is to mix the excited states so that this value is usually reduced by small amounts. Some times, the g values are not defined  by the spin flip transitions of H.S but they are obtained by equating the Zeeman energy $g\mu_BH$ to a band gap energy, $\Delta$ in which case $g=\Delta/\mu_BH$  is not related to the free electron value for which ${\it l}$=0 but such values are to be interpreted by using the band structure.

    For $S$=0, there are spin singlets. We consider that one electron has spin up and the other spin down. Substituting S=0 in the above formula, we obtain,
\beg
{1\over 2}g_{\pm}= {1\over 2}
\eeq
for all values of ${\it l}$. For ${\it l}$=0, we obtain 
superconducting states which are similar to ``zero-resistance 
states". For finilte ${\it l}$, there is a cancellation so that
 ${1\over 2}g_{\pm}$=1/2. The 
resistivity at the plateau is $\rho_{xy}=h/ie^2$ which at i=1/2 is $\rho_{xy}=2h/e^2$. Near the plateau the flux is quantized so that the resistivity is zero. Away from the zero-resistivity state where the resistivity is less than $h/e^2$, the Landau levels have a separation given by the cyclotron frequency,
\beg
(n+{1\over 2})\hbar\omega_c=\epsilon_F
\eeq
so that oscillations in the resistivity occur every time the field is changed so that $n$ changes by 1 at a time.
\beg
n={\epsilon_F\over \hbar\omega_c}
\eeq
We make use of the formula which changes the cyclotron frequency to ${1\over 2}g_{\pm}\hbar\omega _c$. At $s$=0, ${1\over 2}g_{\pm}=
{1\over 2}$ so that a factor of half arises in the cyclotron frequency or n changes to n/2. So whereas there were oscillations at 0,1,2, ... etc. Now we have oscillations at 0, 1/2, 1, 3/2, 2, ..., etc. {\it In other words we had oscillations with frequency $\omega_c$ but for
 $s$=0, we have them at $\omega_c/2$. This will double the 
oscillations when $s$=0 is generated}. If both $s$=+1/2 as well as
 -1/2 are populated, $s$=0 may occur and for finite ${\it l}$ we have oscillations in the resistivity, i.e., $s$=0, ${\it l}\ne 0$
 oscillates whereas
$s$=0, ${\it l}$=0 superconducts.

\noindent{\bf3.~~ Experiment}.

     The samples are modulation doped $GaAs/Al_{0.3}Ga_{0.7}As$ heterostructures. The layers were grown by MBE on 001 face of a GaAs wafer. A 40 nm undoped $Al_{0.3}Ga_{0.7}As$ spacer layer was used to separate 50 nm thick Si donor layer with $n=10^{18} cm^{-3}$ from the 2DEG formed between the spacer and a 500 nm GaAs buffer layer. NiCr gates were evaporated on both sides. The ESR is detected by a probe current of 200 nA at 720 Hz. The microwave is modulated with a 
frequency of 10.8 Hz. The change in resistivity due to microwaves is measured
by another amplifier. The details of such measurements are described 
by Jiang et al[11].

     In Fig.1 we show the transverse resistivity, $\rho_{xx}$, as a function of magnetic field. The figure (1a) shows the usual 
Shubnikov-de Haas oscillations due to the oscillations in the response function. At low fields, large values of $n$ are seen. As the field increases, the value of $n$ reduces. The values 4, 5 and 6 are clearly visible. At these magnetic fields only one spin orientation is 
available so that all of the oscillations arise from only one spin value. When microwave power of about 1 mW at the frequency of 13.1 
Hz is turned on, the other spin orientation gets populated and $s=0$ mode condenses so that the cyclotron frequency reduces to ${1\over 2}\omega_c$. Hence the number of oscillations doubles. In figure (1b) 
we show the change in the resistivity as a function of magnetic field with 13.1 GHz microwave turned on. If we count the number of oscillations from 0 to 1 Tesla at least 5 maxima are seen with 
microwave turned off, whereas 10 maxima are seen with microwave turned on. This means that there are two times more oscillations with 
microwave ${\it on}$ than with microwave ${\it off}$. This is the phenomenon born out due to two spin orientations with $S=0$.

     The g-value of -0.44 belongs to the band gap in GaAs. It varies
 due to spin-orbit interaction and due to temperature. At low temperatures, it is closer to the band value. At 14.1 GHz the ESR line occurs at about 2.672 Tesla. This field is higher than shown in figure. The ESR is obviously a different phenomenon than the doubling of lines. In the experiment performed by Jiang et al an electric field is 
applied. For small electric fields, $\sim$ 1000 V/cm the ESR line 
occurs at $\sim$ 0.382 whereas at larger electric fields, $\sim$ 5000 V/cm a different line is seen at -g =0.377. Adding these two values we get 0.759. The first value normalized by the sum is 0.5033 whereas the second value is 0.4967. These values are $0.5\pm 0.0033$. The formula gives $0.5\pm S/(2{\it l}+1)$. These deviations are a result of the spin-orbit interaction and the changes in the band energies due to temperature.
For ${\it l}$=2, s=1/2, the predicted ${1\over2}g_-$=2/5=0.4 so the formula is consistent with the data. Adding the electric field is equivalent to adding eV in the energy so that different direction 
shifts result due to $\pm$ sign in the formula.

\noindent{\bf4.~~ Conclusions}

      At S=0, ${\it l}=0$, the charge per particle is 1/2. Two such particles make one electron. The 1/2 charge gives half the cyclotron frequency and doubles the number of oscillations. These phenomenon are independent of ESR. 

     There is splitting of ESR line due to doubling in the g values
 such that when electric field is applied one line moves towards lower frequency side (red shift) while the other line moves towards higher frequency side (blue shift). There is a quasiparticle of zero charge. The g values occuring at 2.0023 and at -0.44 are found to have quite different origin. The deviations from our formula are a result of spin-orbit interaction.

\vskip1.25cm
\noindent{\bf5.~~References}
\begin{enumerate}
\item R. G. Mani, et al, Nature {\bf419}, 646(2002)
\item R. G. Mani, J. H. Smet, K. von Klitzing, V. Narayanmurti, W. B. Johnson and V. Umansky, cond-mat/0306388; cond-mat/0303034.
\item A. F. Volkov, cond-mat/0302615
\item F. S. Bergeret, B. Huckestein and A. F. Volkov, cond-mat/0303530.
\item K. N. Shrivastava, Phys. Lett. A {\bf 113}, 435 (1986).
\item K. N. Shrivastava, Mod. Phys. Lett. B {\bf 13}, 1087 (1999).
\item K. N. Shrivastava, Mod. Phys. Lett. B {\bf 14}, 1009 (2000).
\item K. N. Shrivastava, cond-mat/0303309, cond-mat/0303621, cond-mat/0212552.
\item K. N. Shrivastava, Superconductivity: Elementary Topics,
World Scientific, New Jersey, London, Singapore, 2000.
\item K.N. Shrivastava, Introduction to quantum Hall effect,\\ 
      Nova Science Pub. Inc., N. Y. (2002).
\item H. W. Jiang and E. Yablonovitch, Phys. Rev. B {\bf 64}, 041309 (2001).
\end{enumerate}
\vskip0.1cm

Note: Ref.5 is available from:\\
 Nova Science Publishers, Inc.,\\
400 Oser Avenue, Suite 1600,\\
 Hauppauge, N. Y.. 11788-3619,\\
Tel.(631)-231-7269, Fax: (631)-231-8175,\\
 ISBN 1-59033-419-1 US$\$69$.\\
E-mail: novascience@Earthlink.net

Fig.1. (a) Upper curve shows the diagonal resistivity as a function of magnetic field showing Shubnikov-de Haas oscillations. The Landau level
quantum numbers 6, 5, and 4 are marked as a function of increasing magnetic field. The microwave is off. (b) Lower graph shows the change in resistivity upon turning on the 13.1 GHz microwaves. Due to condensation at $\omega_c/2$ the number of oscillations doubles, so that lower graph has twice the number of oscillations than the upper graph.

\end{document}